\documentclass[lnbip]{svmultln}

\newcommand{\orcidLM}	{\href{https://orcid.org/0000-0002-6866-0799}{\protect\includegraphics[scale=0.045]{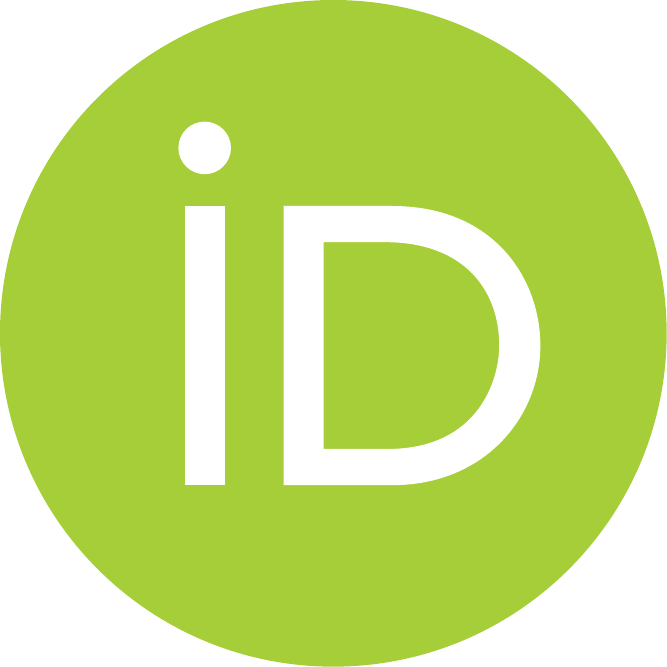}}}

\newcommand{\orcidJG}	{\href{https://orcid.org/0000-0002-7932-5822}{\protect\includegraphics[scale=0.045]{img/orcid}}}

\newcommand{\orcidRB}	{\href{https://orcid.org/0000-0002-5515-7158}{\protect\includegraphics[scale=0.045]{img/orcid}}}

\usepackage{makeidx}  
\usepackage{graphicx} 
\usepackage{booktabs} 
\usepackage{xparse}   
\NewDocumentCommand{\rot}{O{35} O{1em} m}{\makebox[#2][l]{\rotatebox{#1}{#3}}}%
\newcommand{\eg}{e.\,g.,~}
\newcommand{\wrt}{w.\,r.\,t.~}
\newcommand{\ie}{i.\,e.,~}
\usepackage[USenglish]{babel}
\addto\captionsUSenglish{}
\usepackage{lmodern}
\usepackage{adjustbox}
\usepackage{tabto}
\usepackage{booktabs}
\usepackage{tabularx}
\usepackage{makecell}
\usepackage{rotating}
\usepackage{graphicx}
\usepackage{tikz}
\usepackage{pgfplots}
\pgfplotsset{compat=1.15}
\usepackage{multicol,multirow}
\usepackage{amssymb,amsmath}
\usepackage[autostyle, english=american]{csquotes}
\usepackage{url}
\usepackage{xcolor}
\usepackage{placeins}
\usepackage[inline, shortlabels]{enumitem}
\usepackage[locale=US]{siunitx}
\usepackage{subcaption}
\usepackage{todonotes}

\usepackage{listings}

\usepackage{hyperref}

\usepackage{listings}
\usepackage{tabto}
\usepackage{booktabs}
\usepackage{tabularx}
\usepackage{subcaption}

\begin{document}

\makeatletter
\def\RemoveSpaces#1{\zap@space#1 \@empty}
\makeatother

\renewcommand{\thefigure}{\arabic{figure}}
\mainmatter              
\title{An IoT-Enriched Event Log for Process Mining in Smart Factories}

\titlerunning{IoT-Enriched Event Log for Process Mining in Smart Factories}  
%
\author{Lukas Malburg\inst{1,2}\orcidLM \and 
Joscha Grüger\inst{1,2}\orcidJG \and 
Ralph Bergmann\inst{1,2}\orcidRB}
\authorrunning{L. Malburg et al.}   
%
\tocauthor{Lukas Malburg, Joscha Grüger, Ralph Bergmann}
\institute{Artificial Intelligence and Intelligent Information Systems, \\ University of Trier, 54296 Trier, Germany\\ \email{{malburgl,grueger,bergmann}@uni-trier.de}\\
\url{http://www.wi2.uni-trier.de} \and German Research Center for Artificial Intelligence (DFKI) \\ Branch University of Trier, 54296 Trier, Germany\\ \texttt{\{lukas.malburg,joscha.grueger,ralph.bergmann\}@dfki.de}}

\maketitle              

\begin{abstract}
Modern technologies such as the \emph{Internet of Things (IoT)} are becoming increasingly important in various domains, including \emph{Business Process Management (BPM)} research. One main research area in BPM is process mining, which can be used to analyze event logs, \eg for checking the conformance of running processes. However, there are only a few IoT-based event logs available for research purposes. Some of them are artificially generated and the problem occurs that they do not always completely reflect the actual physical properties of smart environments. In this paper, we present an IoT-enriched XES event log that is generated by a physical smart factory. For this purpose, we create the SensorStream XES extension for representing IoT-data in event logs. Finally, we present some preliminary analysis and properties of the log.    

\end{abstract}
\section{Introduction}
\label{sec:Introduction}

The combination of \emph{Business Process Management (BPM)} methods with the \emph{Internet of Things (IoT)} promises several advantages for both sides \cite{Janiesch.2020_Manifesto}. The smart environment sensed and actuated by IoT-devices can benefit from process modeling methods for controlling data acquisition and actuation of resource functionalities \cite{Seiger.2022_IntegratingProcessManagement,Malburg.2020_FactoriesAndBPM}. Moreover, \emph{Process Mining (PM)} \cite{ProcessMiningManifesto} techniques can be applied in smart environments \cite{Leotta.2015_ProcessMining} such as manufacturing \cite{RinderleMa.2021_ProcessAutomationAndMining,Seiger.2022_IntegratingProcessManagement} to check conformance \wrt the given process model or to adapt and optimize processes when runtime failures occur. On the other hand, BPM can benefit from systematic data collection and the variety of IoT-data, \eg event data or context data. BPM research artifacts can then be modified to achieve appropriate analysis results with this more complex IoT-data. In current research, some event logs for PM in smart environments have been proposed (\eg \cite{stertz2020analyzing,Zisgen.2022_IoTEventLogGenerator,Serral.2022_SupportingUsers}). Even though these provide a good basis for research, they are mostly synthetically generated, as acquiring real-world data for research purposes can be very difficult \cite{Malburg.2020_FactoriesAndBPM,Klein.2019_PredM}. However, artificially generated data does not always completely reflect the actual physical properties of smart environments, such as runtime behavior and ad-hoc interventions \cite{Malburg.2020_FactoriesAndBPM}. Another current issue is the lack of support by the \emph{eXtensible Event Stream (XES)} format~\cite{gunther2014xes} to represent IoT-data appropriately in the log. In previous work \cite{Grueger.2022_SensorStreamXESExtension}, we introduce a first step towards a XES extension to enable representing IoT-enriched event logs. In this paper, we present a concrete IoT-enriched event log produced by our physical smart factory. The log is available at \cite{Malburg.2022_EnrichedEventLogDataset} with documentation. The main advantage by using small-scale physical simulation models is that it enables to conduct laboratory experiments while maintaining real world environmental conditions of production lines. Thus, they provide much more realistic data than synthetically generated data. A further advantage is that developed research artifacts can be evaluated by using the factory model. This strengthens the results in a scenario that is closer to real-world production lines and, thus, facilitates the transfer to them \cite{Malburg.2020_FactoriesAndBPM,Seiger.2022_IntegratingProcessManagement}. 

In the following, Sect.~\ref{sec:DataAcquisition} describes how the proposed IoT-enriched event log has been generated in the physical smart factory. Afterwards, a detailed description of the event log is given in Sect.~\ref{sec:descdataset}. Preliminary analyses are presented in Sect.~\ref{sec:prelimanal} and, finally, Sect.~\ref{sec:Conclusion} summarizes the paper.

\section{Data Acquisition}
\label{sec:DataAcquisition}

\begin{figure}[!htb]
  \centering
  \includegraphics[width=0.6\textwidth]{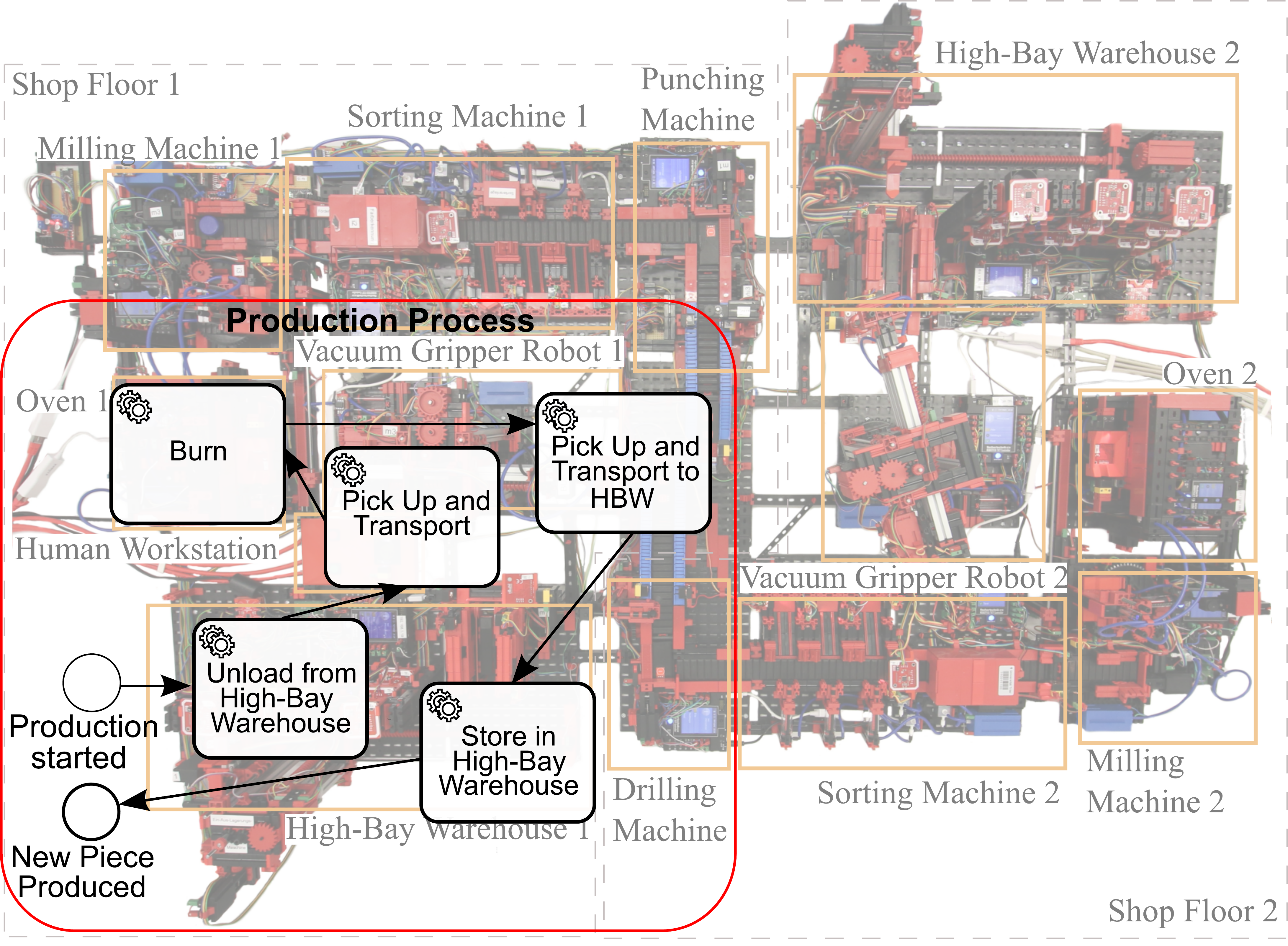}
  \caption{Process-based Control of the Fischertechnik Factory Simulation Models. \cite{Seiger.2022_IntegratingProcessManagement}}
  \label{img:grafik-factoryModel}
\end{figure}

\begin{figure}[htb]
\centering
  	{\includegraphics[width=1.0\textwidth]{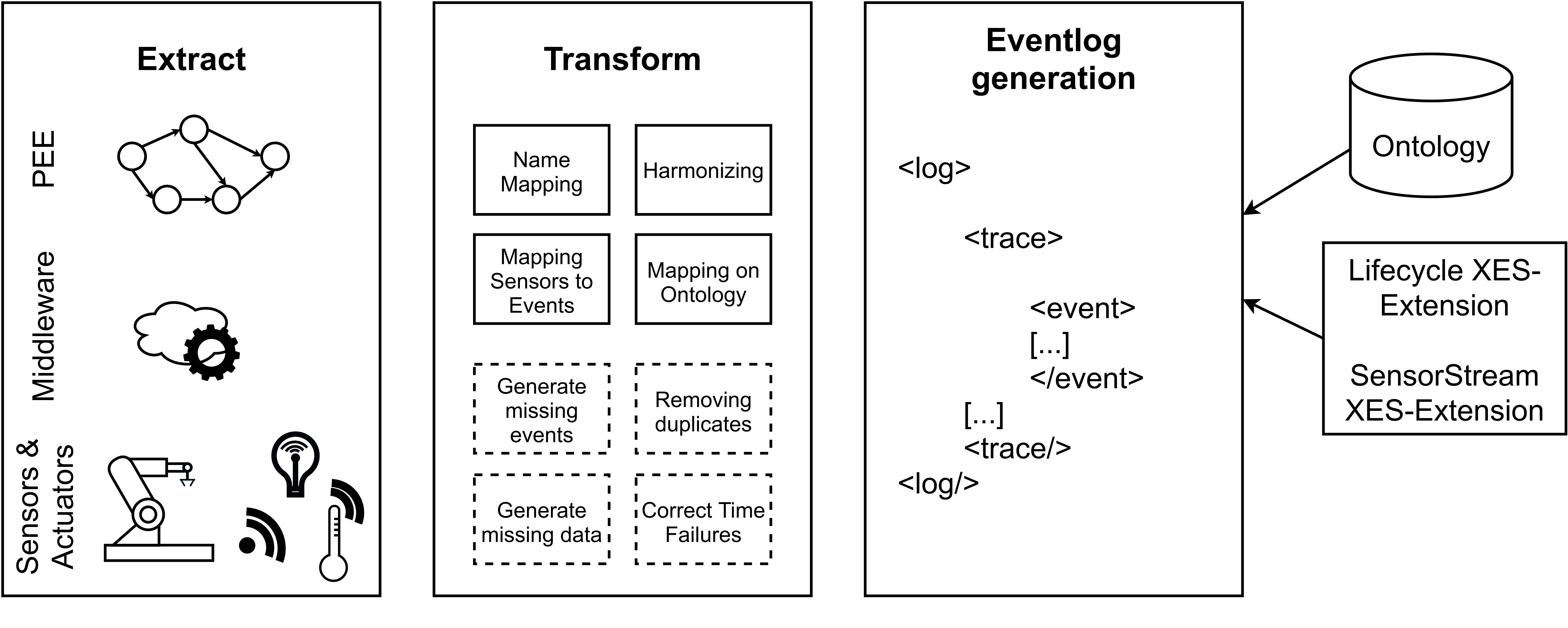}}
	\caption{Pre-processing of data From the Process Execution Engine (PEE), the middleware and from the actuators and sensors. The dashed lines indicate which steps were applied only to the corrected log. \label{fig:PreProcessing}}
\centering
\end{figure}

To generate our proposed IoT-enriched event log, we use a \emph{Fischertechnik (FT)} physical factory simulation model illustrated in Fig~\ref{img:grafik-factoryModel}\footnote{\url{https://iot.uni-trier.de}}. The factory represents two production lines that work independently of each other but are connected for the exchange of workpieces. 
Each shop floor consists of six identical machines. In addition, there are individual machines on each shop floor, \ie a \emph{Punching Machine (PM)} and a \emph{Human Workstation (HW)} on the first shop floor and a \emph{Drilling Machine (DM)} on the second one. To enable BPM-related research, we use a service-based architecture \cite{Seiger.2022_IntegratingProcessManagement,Malburg.2020_SemanticWebServices_IN4PL,Malburg.2020_FactoriesAndBPM} to control the production lines in a process-based fashion by \emph{Workflow Management Systems (WfMSs)}. The data produced by the machines is pushed to Apache Kafka\footnote{\url{https://kafka.apache.org/}}. Based on this endpoint, we extract the IoT-data from Kafka and the data on process executions from the used web server, which contains further concrete information compared to the log from the WfMS. For example, besides the start and end time of each activity, the reasons for runtime exceptions and the planned operation times are also logged (see Sect.~\ref{sec:descdataset}). We model 16 processes that are executed in the smart factory. In total, we execute processes in the factory for more than 20 hours and recorded its IoT-data consisting of data from sensors and actuators as well as the process data. Based on the three data sources, we build the IoT-enriched event log by applying a comprehensive pre-processing. We publish the log in two versions, an original version in which harmonization of data, unification of names, mapping of sensor data to events from the \emph{Process Execution Engine (PEE)}, and a mapping of events, actuators, and sensors to entities in the underlying FTOnto ontology \cite{Klein.2019b_FTOnto} of the factory were performed (see figure \ref{fig:PreProcessing}). On the other hand, a corrected event log based on the pre-processed original log is proposed. Here, in addition to the basic pre-processing, errors in the data were also corrected. To this end, missing events were restored, duplicates were removed, missing data was generated, and errors caused by time shifts were corrected. Using the SensorStream XES extension, the log was then generated \cite{Grueger.2022_SensorStreamXESExtension}. 
\section{Description of the IoT-Enriched Event Logs}
\label{sec:descdataset}
\begin{figure}[htb]
\centering
  	{\includegraphics[width=.70\textwidth]{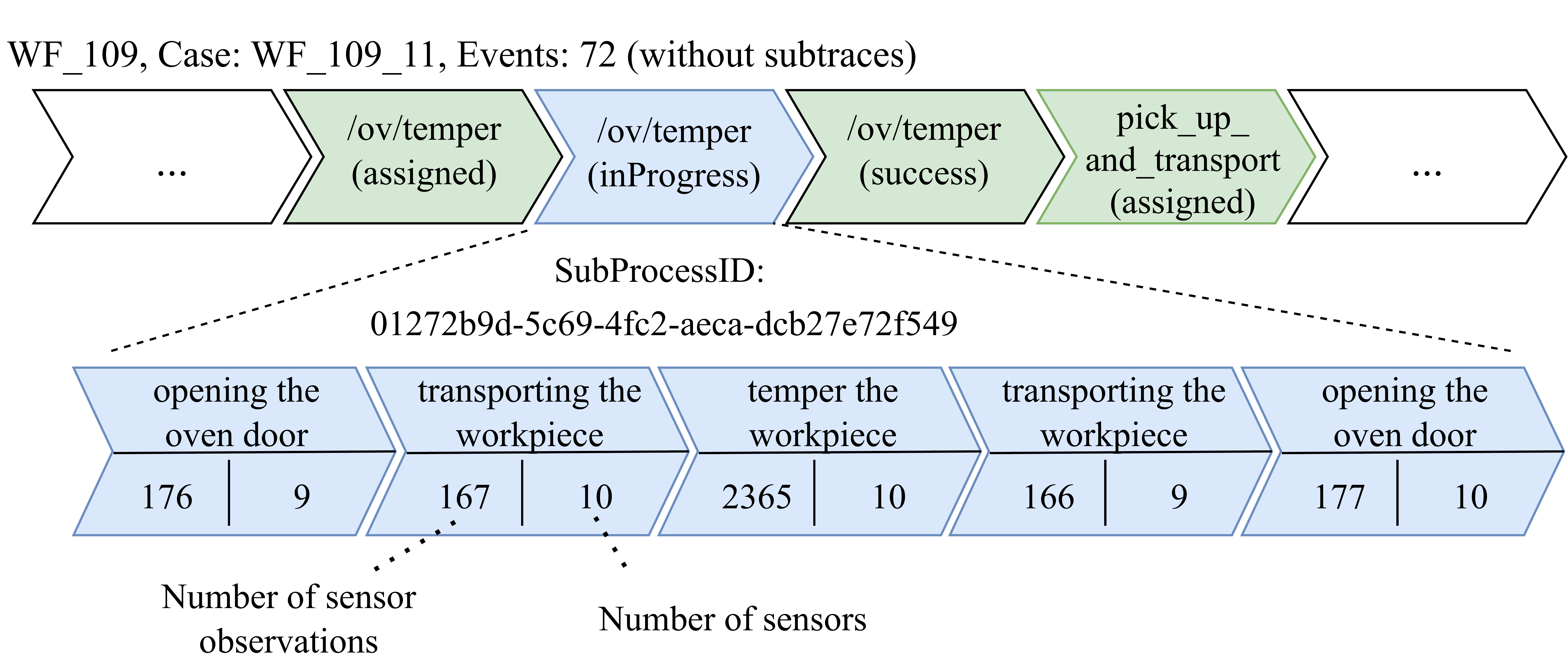}}
	\caption{Excerpt of a Process Instance of Workflow 109 Including Resolving the Relationship to the Subprocess of the Event "/ov/temper" in the Lifecycle State "(Inprogress)" via the SubProcessID. \label{fig:subprocess}}
\centering
\end{figure}

All event logs are provided in XES format. The Concept, Identity, Time, Lifecycle, and the SensorStream XES \cite{Grueger.2022_SensorStreamXESExtension} extensions are used to represent the data. The SensorStream XES extension enriches the XES standard with the ability to integrate complex sensor data into the event log. For this purpose, the extension introduces the \texttt{sensorstream} schema. This uses the SOSA and SNN\footnote{\url{https://www.w3.org/TR/vocab-ssn/}} ontology to semantically describe the sensor data, \eg which sensor produces the data and to which actuator this sensor is attached. In the event logs provided, the extension is used to describe sensor data at the event level, which forms the context of the event execution. For each sensor value the data type, the interaction type (observation or actuation), the type of the receiving system as well as the assignment in the underlying FTOnto \cite{Klein.2019b_FTOnto} ontology, and the timestamp are described. The log reflects several levels of granularity. For this purpose, events refer via the \emph{SubProcessID} attribute to the subtrace, which describes the respective event with finer granularity (see Fig.~\ref{fig:subprocess}).
First, we provide an unmodified and error-ridden IoT-enriched event log. The errors in the event log are typical problems in real-world production shop floors or, in general, in IoT domains: For example, events in the log are delayed or completely missing due to communication problems, or are sent multiple times and logged twice due to timeouts. In addition, the response code of the web server is not contained in some events. For this reason, it is not possible to obtain whether the corresponding activity was successfully executed or not in the shop floor. The timestamps of IoT-sensor data and the corresponding process data are sometimes shifted since a Network Time Protocol server was out of service. Second, we provide the event log in a cleaned gold standard. Therefore, all errors and \emph{Data Quality Issues (DQIs)} were removed \cite{Bose2013-sb}. Using our developed domain ontology FTOnto \cite{Klein.2019b_FTOnto} and separate recordings produced by Apache Kafka from the smart factory used for logging, it was possible to correct the timestamps and to reconstruct unrecorded, so-called invisible events. By providing these comprehensive IoT-enriched event logs, we enable process mining approaches to be developed and evaluated based on real-world data. In particular, the provision of a real version of the dataset interspersed with DQIs and a cleaned dataset enables the explicit targeting of DQIs. Thus, we provide the data for addressing the process mining challenges C1) Finding, Merging, and Cleaning Event Data, C2) Dealing with Complex Event Logs Having Diverse Characteristics, and C4) Dealing with Concept Drifts \cite{ProcessMiningManifesto}.

\section{Preliminary Analysis}
\label{sec:prelimanal}
In the preliminary analysis, we analyze the event logs, their structure, and content. Unless otherwise stated, the values refer to the original and erroneous event log (see Tab.~\ref{tab:OverviewDatasets}). Sensors are only present at the resource level, and sensor values are directly associated with events in the subtraces. The number of sensors and sensor values varies between resources. While \emph{High Bay Warehouse (HBW)} sometimes records 160 sensor values per second or more, \emph{Human Workstation (HW)} only records about 40 values. 

In total, 16 workflows are shown. While on the \emph{MainProcess} level (Main log in Tab. \ref{tab:OverviewDatasets}), all events represent the three lifecycle steps \emph{assigned}, \emph{inProgress}, and \emph{success} or \emph{failure}. In the sublog (sublogs in Tab. \ref{tab:OverviewDatasets}), no lifecycle attributes are present. All traces start with a \emph{/hbw/unload}. The end event is \emph{/hbw/store} in 212 traces and \emph{/hbw/unload} in 34 traces. The end of the other traces is distributed over 14 further events. A closer look at workflow 111 shows that each event has an edge to itself. These represent the events in different lifecycle stages. It also shows that the traces can end in any of the activities. This suggests that the process of burning the workpiece in the oven, milling it in the milling machine, and then deburring it in the milling machine contains errors. 

Since the SensorStream extension has not yet been implemented in any process mining tool, the sensor perspective could not be analyzed. Due to a lack of space, the illustration of a process discovery was omitted.

\begin{table}[]
    \centering
    \caption{Overview Over the Logs (after preprocessing, corrected log). Traces in the main log reference on sublogs.}
\begin{tabular}{lccccccccm{0.5cm}m{0.5cm}m{0.5cm}}
 \textbf{Data Sets}   
    & \textbf{\rot{Events}}
    & \textbf{\rot{Cases}}
    & \textbf{\rot{Activities}}
    & \textbf{\rot{Resources}}
    & \textbf{\rot{Variants}}
    & \textbf{\rot{Actuators}}
    & \textbf{\rot{Sensors}}
    & \textbf{\rot{Data Points}}
    & \multicolumn{3}{c}{\textbf{\makecell{Trace Len\\(avg,min,max)}}}\\
    
    \midrule
    \textbf{Main log} &  9,471 & 301 & 21 & 15 & 231 & - &- & - & 31 & 3 & 69\\
    \hline
     \textbf{Sublogs} &  13,424 & 3,118 & 109 & 15 & 269 & 52 & 131 & 136,208,108 & 8 & 1 & 14\\
     \hline
     \textbf{Total} &  \textbf{22,895} & \textbf{3,489} & \textbf{21} & \textbf{15} & \textbf{500} & \textbf{52} & \textbf{131} & \textbf{136,208,108} & \textbf{10} & 1 & \textbf{69}\\
\end{tabular}
    \label{tab:OverviewDatasets}
\end{table}

\section{Conclusion}
\label{sec:Conclusion}
In this paper, we present an IoT-enriched event log for process mining research in smart factories. Such event logs are currently rather rare and mostly synthetically generated. We generate two versions of the event log with a physical factory model (see Sect.~\ref{sec:DataAcquisition}): one version is the native log with DQIs such as delayed or missing events etc. and a cleaned version without these DQIs. The IoT-enriched event logs provide the basis for data-intensive research involving IoT sensor data in the process mining field. However, the current frameworks and tools for process mining cannot directly use the additional data. In future work, we investigate to extend frameworks and tools to facilitate this. 

\FloatBarrier
\interlinepenalty10000
\bibliographystyle{splncs}
\bibliography{bibliography}

\end{document}